\documentclass[preprint,showpacs,preprintnumbers,amsmath,amssymb]{revtex4}

\usepackage{graphicx}
\usepackage{dcolumn}
\usepackage{bm}

\begin{document}

\preprint{APS/123-QED}

\title{A braid model for the particle X(3872)\\}

\author{C.  Pe\~na}
\email[carlos.andres.pena.castaneda@pwr.wroc.pl]{}
\author{L. Jacak}
\affiliation{%
Institute of Physics, Wroc{\l}aw University of Technology,
Wyb. Wyspia{\'n}skiego 27, 50-370 Wroc{\l}aw, Poland
}%

\date{\today}

\begin{abstract}
The Model of Quark Exchange (MQE) describes the particle X(3872) as a meson molecule. 
We asked whether braids influence the meson potential in the MQE.
We used the Burau representation that parameterized braids 
with a variable $t$. The present result shows that $t$ rescales the coupling of the meson potential 
determining if it is attractive or repulsive.  As a consequence, a capture diagram favored the molecular state  for   
$t=0.85$,  it breaks for other values. For the future, braids may help to study others exotic states
in geometrical terms.
\end{abstract}

\pacs{Valid PACS appear here}

\pacs{14.20.Pt, 03.65.Nk, 03.65.Fd}
\keywords{Quantum Hall effects, braid groups, knot theory}

\maketitle

\section{\label{sec:level1}Introduction}
In 2003  the  Belle  Collaboration discovered the particle X(3872) \cite{Belle2003} which other experiments confirmed  
later \cite{CDF2004, DO2004, BABAR2005, LHCb2012}. More  recently, the detector LHCb measured its 
quantum numbers [2].  These findings suggested that X(3872) is composed of something more than two heavy charm quarks ($\bar{c}c$). 
Nowadays, its composition includes one of the alternatives (e.g. \cite{Hambrock2013}):  a molecule made of two heavy mesons 
\cite{Tornqvist2004, Gamermann:2009uq}; a tetra quark \cite{Maiani2005} or a hybrid state.

The Model of Quark Exchange (MQE) describes  the particle X(3872) as a molecule composed of  
mesons $D^0$ and $\bar{D}^{0*}$ (D-mesons) [4]. The model request a potential energy characterized  by a coupling $\lambda_{ex}$ and a width $\gamma$ that were fitted with the 
mass of the X(3872). Moreover, the nature of these parameters was well established in terms of  diagrams called 
capture and transfer \cite{Barnes1992, Martins:1994hd}. Although, the diagrams posses  a vertex signalizing the exchange of two quarks, it is 
unclear how the vertex contributes to the potential energy. We answered this question assuming the capture and transfer 
diagrams content braids. Braids form groups of great interest in physics as they  explain the origin of fermion statistics  
associated with exchange of particles [5]. We proved  that braids modified the phase-shift calculated previously 
in ref. [4] because the coupling $\lambda_{ex}$ rescales by a factor originated from  braids.

The section~\ref{sec:level5}  describes briefly the method based on the quark exchange model for the particle X(3872).
In the section~\ref{sec:level6} we included the generators of the braid group ${\cal B}_3$ in the quark exchange model.

\section{\label{sec:level5} A braid model for the  $X(3872)$}

\subsection{\label{sec:level5.1}The phase-shift}

The phase-shift ($\delta$) describes the state X(3872)  when solving the 
Lippmann-Schwinger equation for the scattering process $D^0+\bar{D}^{0*}\to J/\psi+\rho$ \cite{CarlosX}. 
The authors in ref. \cite{CarlosX} implemented a Lorentzian potential 
for mesons characterized by two parameters: Its strength $\lambda=20.3$~GeV$^{-2}$ and a cutoff $\gamma=0.8$. 
For these parameters the state X(3872) materializes as a molecule of D-mesons with mass of $3.872$~GeV.
The energy variable $z$ enters in the scattering phase-shift $\delta(z)$ defined as
\begin{eqnarray}\label{phaseshift}
\delta(z) &=& \arctan\left(\frac{\textnormal{Im}[\textnormal{t}(z)]}{\textnormal{Re}[\textnormal{t}(z)]}\right)~,
\end{eqnarray}
where  $\textnormal{t}(z)$  relates to the T-matrix that solves the Lippmann-Schwinger equation (see eq.~10 of ref.~\cite{CarlosX}).

\subsection{\label{sec:level5.2}The quark exchange mechanism}
The MQE proposed that a meson potential
originates when the two D-mesons exchange two quarks \cite{barnes:1992, barnes:2001, Martins:1994hd, Blaschke:1992qa}.
This description supposes that all heavy mesons interact with the potential
\begin{eqnarray}\label{potential1App3}
 \textrm{\textbf{\emph{U}}}(p,p') &=&-\textit{C}_{SFC}\;I(p,p')
\end{eqnarray}
with the factor $C_{SFC}=\frac{1}{6}$ and $I(p,p')$ as the invariant matrix element 
of the scattering $D^0+\bar{D}^{0*}\to J/\psi+\rho$. This matrix element  
was calculated using four contributions named capture and transfer diagrams \cite{Martins:1994hd}.
The first capture diagram ($D1$) is shown in  Fig.2.
\begin{figure}[!htbp]
         \includegraphics[height=2.5in]{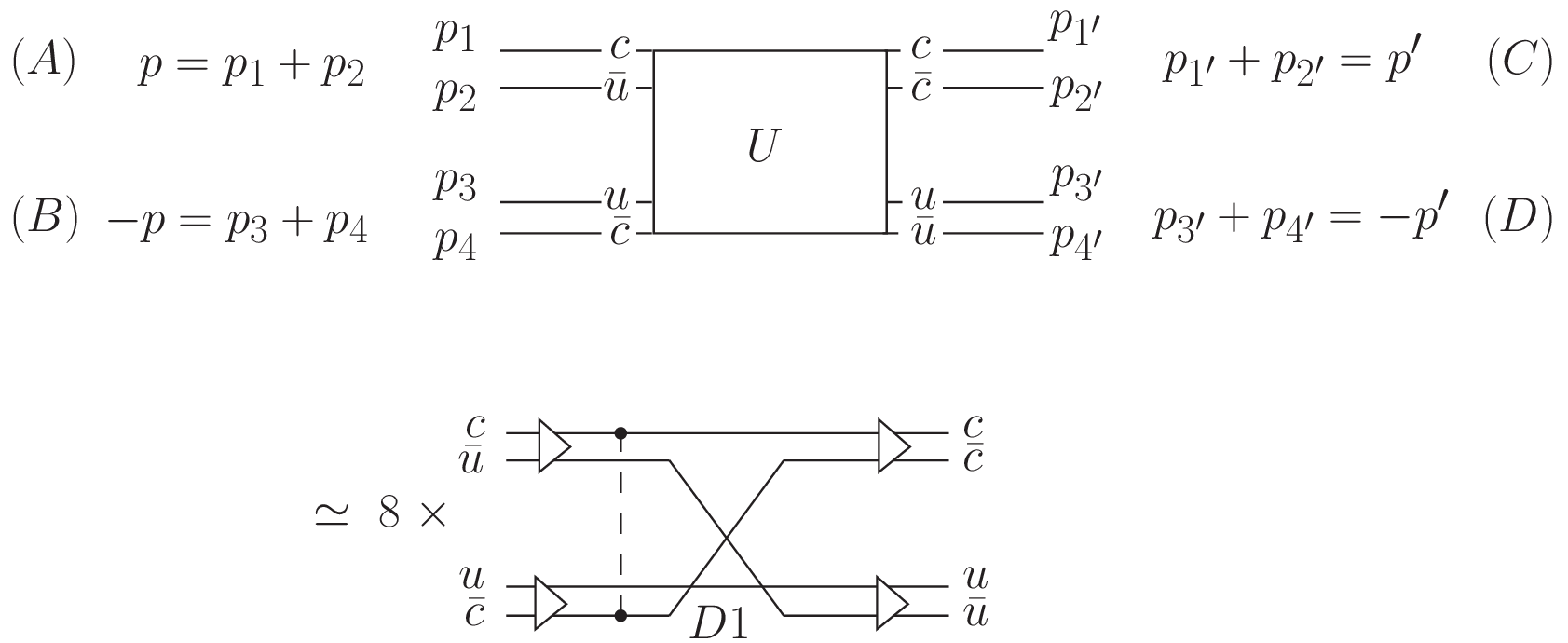}
    \caption{Up panel: A box diagram describes the meson potential.  
D-mesons are composed of the quarks up (u) and charm (c).
Down panel: The meson potential approximates to the capture diagram $D1$ in the
first ladder approximation. The factor eight derives from the product of:
the two conjugate channels $D^0$, $\bar{D}^{0*}$ and $\bar{D}^0$, $D^{0*}$ \cite{Martins:1994hd};  
the  prior and post interaction  \cite{Martins:1994hd,Blaschke:1992qa}; 
the initial and final momenta of each heavy quark also contributes with a factor two.}\label{figU}
\end{figure}
The result for the capture diagram $D1$ \cite{Martins:1994hd} was 
\begin{eqnarray}\label{matrix1App3}
  I(p,p') &=& \;\langle \phi_{p}\;\phi_{-p}|-\textbf{D}_1|\phi_{p'}\;\phi_{-p'} \rangle
   \nonumber\\
       &=& \displaystyle\sum\limits_{p_1...p_{4^{'}}}\underbrace{\phi_{p}^*(p_1,p_2)}_{A}\underbrace{\phi_{-p}^*(p_3,p_4)}_{B}\times
  \nonumber\\
  &&\bold{V}(p_1,p_4)\delta_{p_2,p_{4^{'}}}\delta_{p_3,p_{3^{'}}}
\nonumber\\
  &&\times \underbrace{\phi_{p'}(p_{1^{'}},p_{2^{'}})}_{C}\underbrace{\phi_{-p'}(p_{3^{'}},p_{4^{'}})}_{D}~.
   \end{eqnarray}
within the quark-quark interaction  
 \begin{equation}\label{expotential}
\bold{V}(p_1,p_4)=-V\;e^{-2 (p_1-p_{1^{'}})^{2}}~\delta_{p_{4},p_{2^{'}}+(p_{1^{'}}-p_1)}
 \end{equation}
and $V=113.39$~GeV$^{-2}$ as a parameter  used
to fit the meson spectrum \cite{Martins:1994hd}. In this paper, the volume in momentum space is one and the spin-spin interaction
is neglected since it is hundred times smaller than $V$. The wave function for a meson $A$ is represented by $\phi_p(p_1,p_2)$ 
with amplitude $\phi_A$, similarly for the other mesons. They define the product $\frac{\phi_A \phi_B \phi_C \phi_D}{(2\pi)^6}= 6.312$~GeV$^{-6}$.
In the continuum limit for the invariant matrix, the sum \eqref{matrix1App3} becomes an integration yielding 
the factor $0.12~\text{GeV}^6$. Therefore the capture diagram $D1$ 
contributes with the coupling value \cite{Martins:1994hd,CarlosThesis2013}
\begin{eqnarray}\label{labmdaex}
\lambda_{ex}&=& -8\, C_{SFC}\, \frac{\phi_A \phi_B \phi_C \phi_D}{(2\pi)^6}\, V\, \left(0.12~\text{GeV}^6\right) ~,
\nonumber\\
&=&-114.35~ \text{GeV}^{-2}~.
 \end{eqnarray}
Thus the strength $|\lambda_{ex}|$  is around six times bigger than the coupling of the 
meson potential $\lambda$. Besides the first capture
diagram $D1$ supplies a repulsive interaction. However, the four capture and transfer diagrams
previously introduced in ref.~\cite{Martins:1994hd} reduces the strength until half of $\lambda$ \cite{CarlosThesis2013}.
Probably higher orders in the ladder approximation may help to reduce this disagreement.
However, We solved this problem using braids inside the capture diagram $D1$. 

The phase-shift is calculated using the method previously described in ref.~\cite{CarlosX}
with a capture diagram (Fig.~\ref{figU}) modified by  braids (Fig.~\ref{figB4}).   

\subsection{\label{sec:level1}Definition of  braids}
\begin{figure}[!htbp]
\includegraphics[height=0.9in]{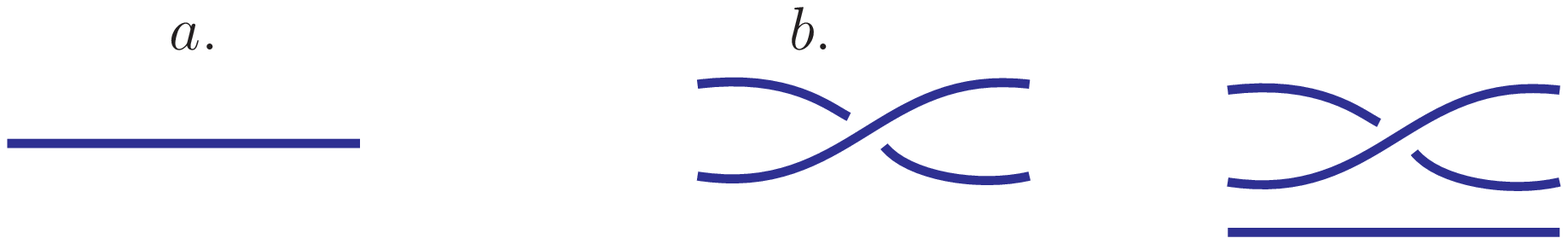}
\caption{\label{fig:braids} Braids as generators of the full braid groups:  a. ${\cal B}_1$,  b. ${\cal B}_2$, c. ${\cal B}_3$.}
\end{figure}

 K. Murasugi  introduced braids in three dimensions as the pictures
in Fig.~\ref{fig:braids} \cite{MurasugiBraidsBook, Murasugi1996}. 
Braids may form groups \cite{Artin1947,TuraevBook2008}. The full braid group  ${\cal B}_n$  
consists of the $n-1$ generators $\sigma_1$, $\sigma_2$, ... $\sigma_{n-1}$ \cite{Artin1947,TuraevBook2008}.
For instance, the group ${\cal B}_3$ is composed of two generators as in Fig.~\ref{fig:braids}c while the groups with one generator
are ${\cal B}_1$ in Fig.~\ref{fig:braids}a (the trivial group) and ${\cal B}_2$ in Fig.~\ref{fig:braids}b (the cyclic group) 
\cite{TuraevBook2008}. The group ${\cal B}_n$ (also called the Artin group) is non-abelian for $n\geq3$. Besides it  emerges as the 
fundamental group of 
$\Re^2$ \cite{Birman1974, Jacak2012}. 

\subsection{\label{sec:level3}The Burau's representation for braids}
W. Burau parametrized the braid group  ${\cal B}_n$  with a variable $t$ 
using an  injective mapping (only for $n<4$) \cite{Burau1935}. He associated to the
element $\sigma_i$ in ${\cal B}_n$ the $n\times n$ matrix 
\begin{equation}\label{psi_n}
\varphi_n(\sigma_i) =
 \begin{pmatrix}
  \begin{tabular}{ l | c |r }
    $I_{i-1}$ &        &  \\ \hline
              & $1-t$ \hspace{0.3cm}   $t$ & \\ 
              & \hspace{0.2cm}1    \hspace{0.6cm}    0 &    \\ \hline
              &        &   $I_{n-i-1}$ \\ 
  \end{tabular} 
 \end{pmatrix}
\end{equation}
\text{with}  $i=1,2,...,n-1$. The empty spaces in \eqref{psi_n} consist of zeros; 
the identity $m\times m$ matrix named $I_m$  disappears from $\psi_n(\sigma_1)$ as $i=1$ and
from $\psi_n(\sigma_{n-1})$ as $i=n-1$. Thus the braid group ${\cal B}_2$ is generated by
\begin{equation}
\varphi_2(\sigma_1) =
 \begin{pmatrix}
1-t & t  \\
  1 & 0 \\
 \end{pmatrix}~.
\end{equation}
and the braid group ${\cal B}_3$ by
\begin{equation}\label{BurauRepresentation}
\varphi_3(\sigma_1) =
 \begin{pmatrix}
  1-t & t & 0 \\
  1 & 0 & 0 \\
  0  & 0  & 1 \\
 \end{pmatrix}~,
\hspace{0.5cm} \varphi_3(\sigma_2) =
 \begin{pmatrix}
  1 & 0 & 0 \\
  0 & 1-t & t \\
  0  & 1  & 0 \\
 \end{pmatrix}~.
\end{equation}
These matrices \eqref{BurauRepresentation} depend uniquely on $t$ in comparison  with other 
representations that use more than one variable \cite{TuraevBook2008, Birman2004, Jackson2004, Rolfsen2010}. 

\section{\label{sec:level6}The Results}
 
\subsection{\label{sec:level8}A braid structure in the  quark exchange diagrams}
\begin{figure}[!htbp]
         \includegraphics[height=1in]{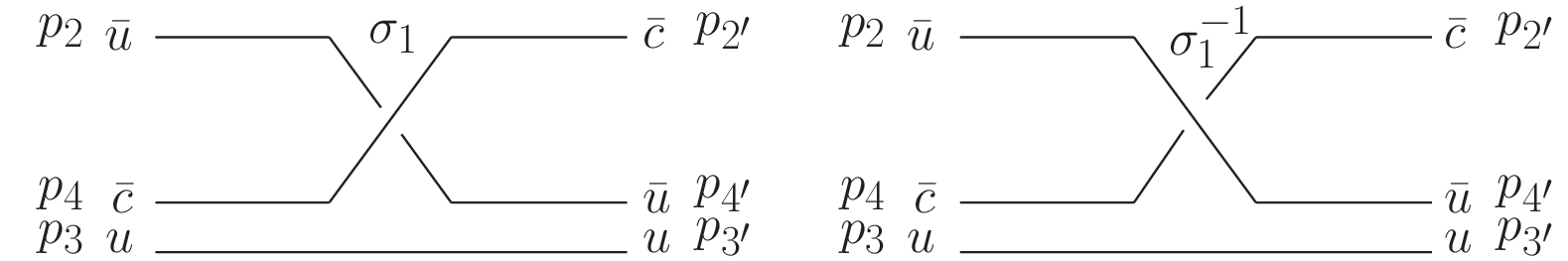}
    \caption{The  braids $\sigma_1$ or $\sigma^{-1}_1$ represents
 the exchange of quarks $\bar{c}$ and $\bar{u}$ as well as in Fig.~\ref{figU}.}\label{figB3}
\end{figure}

\begin{figure}[!htbp]
         \includegraphics[height=5in, angle =-90]{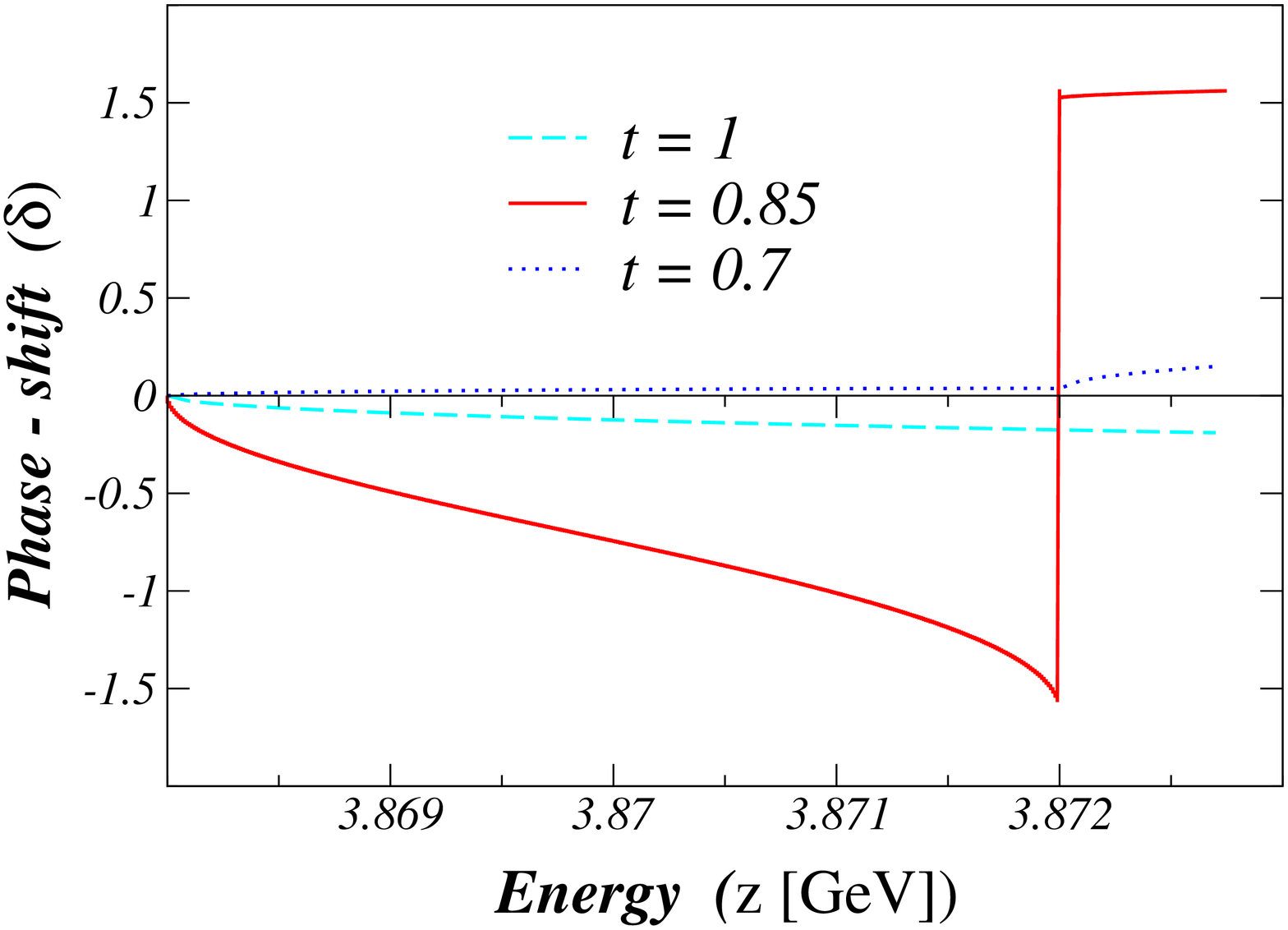}
    \caption{Phase-shift as a function of energy for different values for $t$.  
The phase-shift jumps in $\pi$ at the energy of $3.872$~GeV ($\lambda=20.3$~GeV$^{-2}$, $\gamma=0.8$).}\label{figB4}
\end{figure}

The quark exchange potential \eqref{expotential} determines the capture diagram in 
Fig.~\ref{figU} when the quarks $\bar{u}$ and $\bar{c}$ exchange because of
the constraint given by 
\begin{equation}\label{delta}
 \delta_{p_{4},p_{2^{'}}+(p_{1^{'}}-p_1)}\delta_{p_2,p_{4^{'}}}\delta_{p_3,p_{3^{'}}}~,
\end{equation}
this constraint  permutes the momenta subscripts $(2,3,4)$ 
if the symbol prime is ignored.
The permutation of three numbers is the  element $\Pi_2$ of the permutation group $S_3$ (\cite{Bronshtein2007}, p. 301) 
\begin{equation}\label{permutation1}
\Pi_2= \left( \begin{array}{ccc}
2 & 3 & 4 \\
4 & 3 & 2  
\end{array} \right)  ~.
\end{equation}
Since $\Pi_2$ represents also the $3\times 3$ matrix (\cite{Bronshtein2007}, p. 301-304)
\begin{equation}\label{permutationMatrix}
 \bold{D}\left(\Pi_2\right)=\left( \begin{array}{ccc}
0 & 0 & 1 \\
0 & 1 & 0  \\
1 & 0 & 0
\end{array} \right)~,
\end{equation}
the trace of the permutation matrix \eqref{permutationMatrix}
contributes to the sum \eqref{matrix1App3}. 
This procedure reaches only until fifty porcent of 
the coupling $\lambda$ when including
the whole capture and transfer diagrams together \cite{CarlosThesis2013}.
Besides, the permutation matrix \eqref{permutationMatrix} allows arbitrarily all type 
of quark exchange contributions. We limited the contributions by considering
the exchange diagrams as braids. Each
braid has two alternatives for crossings that distinguishes the type of quark exchange. 
We proved this statement by tracing the product of the permutation
matrix \eqref{permutationMatrix}  with  $\varphi(\sigma_1^{-1})$ and
$\varphi(\sigma_1)$ \eqref{BurauRepresentation}. Hence the traces   
\begin{eqnarray}
 Tr\left[\varphi_3(\sigma_1)\bold{D}(\Pi_2)\right]&=& 0~,\label{multiply1}\\
Tr\left[\varphi_3(\sigma^{-1}_1)\bold{D}(\Pi_2)\right]&=& 1-\frac{1}{t}~.\label{multiply2}
\end{eqnarray}
affect the value of $\lambda_{ex}$ such that  the potential  scales with the new strength
\begin{eqnarray}\label{Rescale}
\lambda &=& \lambda_{ex}\left(1-\frac{1}{t}\right)~, 
\end{eqnarray}
As a result the quarks exchange as the braid $\sigma_1^{-1}$ (Fig.~\ref{figB3}) instead of $\sigma_1$.  
We calculated the phase-shift (\ref{figB4}) as a function of  the energy ($z$) and the variable $t$.
Our method is the same that in ref.~\cite{CarlosX} but with a change induced by
the coupling \eqref{Rescale}. The Lorentzian form factors $L(p)$ and $R(p)$
have cutoff $\gamma = 0.8$. We observed that $t=0.85$
makes a molecule of $D^0$ and $\bar{D}^{0*}$. Moreover,
a sharp behavior of the phase-shift around $3.872$~GeV remains. 
We obtained the same result than in ref.~\cite{CarlosX} only when $t\to \infty$. 

\section{\label{sec:level7}Discussion}
By including braids in the MQE we found that they regulate the repulsive or attractive character of the meson potential \eqref{potential1App3}. 
Besides the braid model explains the origin of the coupling $\lambda$ at the quark level.
Therefore, the braids in the Fig.~\ref{figB3} makes the phase-shift (Fig.~\ref{figB4})
to jump in $\pi$. It confirms that the pair of mesons $D^0$, $\bar{D}^{0*}$  forms a molecule with mass of $3.872$~GeV
for $t=0.85$, other vales destroy it. Although, for the molecule, we obtained zero for the binding energy it may differ
if all capture and transfer diagrams are included \cite{CarlosThesis2013}.

This braid model requires two quarks exchange (one heavy $\bar{c}$ and one light $\bar{u}$) but other models proposed a meson exchange
($\rho$ or $\omega$) containing only light quarks \cite{Gamermann:2009uq, Tornqvist2004, Zhou2014}.
Nevertheless, whether the $X(3872)$ prefers the exchange with two quarks rather than a meson  
is unclear. 

We observed within one capture diagram that  $t=1$
destroys the molecule because the meson potential vanishes (see the phase-shift in Fig.~\ref{figB4}).
Moreover, for $t>1$ the potential gets repulsive (disfavors the molecule) 
while for $t<1$ attractive (favors the molecule). An attractive potential
rises also for $t>1$ when considering all the capture and transfer diagrams leading a small
chance for the braids effects to disappear. 
Although other braids belongs to the braid group ${\cal B}_3$ we have chosen those as
in Fig.~\ref{figB3} with one crossing  so the potential attenuates only with the polynomial $1-\frac{1}{t}$. 
We believe the physical meaning of $t$ would emerge
if a relation between the geometrical and the matrix representation of braids
is established. Unfortunately such relation is absence in literature. 

The robust calculation of the phase-shift in this molecular model 
includes the quark structure of mesons something common in all exotic states. 
The meson potential solves analytically the Lippmann-Schwinger equation since it 
separates in two form factors ($L(p)$, $R(p)$). Either  Lorentzian or Gaussian form factors
yield almost the same phase-shift with small adjustments for the cutoff $\gamma$. 
We neglected the width of the $\rho$ meson 
in the scattering process $D^0+\bar{D}^{0*}\to J/\psi+\rho$
causing the phase-shift to jump sharply at $3.872$~GeV.
A similar behavior is well known for the Deuteron system \cite{Schmidt199057}.

The charged mesons $D^+$ and $\bar{D}^{*-}$ may affect the phase-shift for energies above $3.872$~GeV as previously
was studied in \cite{CarlosThesis2013}. We plan to test this braid model for other exotic states 
since current descriptions do not apply well for all \cite{Hambrock2013}.
The braid model fits well with the predictions for 
a hot and dense medium created  in heavy ion collisions where the X(3872) may form \cite{Carlos2014PRB,Carlos2011PRBProc}.
The capture diagram used in Fig.~\ref{figU} corresponds to the first order
in the ladder approximation. Therefore these findings must be interpreted with caution for 
higher order.

To summarize, braids may help as a mechanism for production of exotic states. 
The braid $\sigma^{-1}_1$ modifies the meson potential
because the coupling $\lambda_{ex}$ rescales by a factor $1-\frac{1}{t}$ such that a $X(3872)$  forms as a molecule of
$D^0$ and $\bar{D}^{0*}$ for $t=0.85$.  The value of $t$ may change significantly  
when adding more capture and transfer diagrams.  Nevertheless, it informs if the meson
potential is attractive or repulsive. Future studies testing this braid model for other exotic states
and predictions for heavy ion collisions are needed.

\begin{acknowledgments}
The support from the NCN Project UMO-2011/02/A/ST3/00116 is acknowledged.
C. Pe\~na thanks the comments of D.~Blaschke.
\end{acknowledgments}

\bibliography{Xparticle}

\begin{thebibliography}{31}
\expandafter\ifx\csname natexlab\endcsname\relax\def\natexlab#1{#1}\fi
\expandafter\ifx\csname bibnamefont\endcsname\relax
  \def\bibnamefont#1{#1}\fi
\expandafter\ifx\csname bibfnamefont\endcsname\relax
  \def\bibfnamefont#1{#1}\fi
\expandafter\ifx\csname citenamefont\endcsname\relax
  \def\citenamefont#1{#1}\fi
\expandafter\ifx\csname url\endcsname\relax
  \def\url#1{\texttt{#1}}\fi
\expandafter\ifx\csname urlprefix\endcsname\relax\def\urlprefix{URL }\fi
\providecommand{\bibinfo}[2]{#2}
\providecommand{\eprint}[2][]{\url{#2}}

\bibitem[{\citenamefont{{S.-K. Choi, \emph{et al.}}}(2003)}]{Belle2003}
\bibinfo{author}{\bibnamefont{{S.-K. Choi, \emph{et al.}}}}
  (\bibinfo{collaboration}{Belle Colaboration}), \bibinfo{journal}{Phys. Rev.
  Lett} \textbf{\bibinfo{volume}{91}}, \bibinfo{pages}{262001}
  (\bibinfo{year}{2003}).

\bibitem[{\citenamefont{{D. Acosta, \emph{et al.} }}(2004)}]{CDF2004}
\bibinfo{author}{\bibnamefont{{D. Acosta, \emph{et al.} }}}
  (\bibinfo{collaboration}{CDF Collaboration}), \bibinfo{journal}{Phys. Rev.
  Lett.} \textbf{\bibinfo{volume}{93}}, \bibinfo{pages}{072001}
  (\bibinfo{year}{2004}).

\bibitem[{\citenamefont{{V. M. Abazov, \emph{et al.}}}(2004)}]{DO2004}
\bibinfo{author}{\bibnamefont{{V. M. Abazov, \emph{et al.}}}}
  (\bibinfo{collaboration}{D0 Collaboration}), \bibinfo{journal}{Phys. Rev.
  Lett.} \textbf{\bibinfo{volume}{93}}, \bibinfo{pages}{162002}
  (\bibinfo{year}{2004}).

\bibitem[{\citenamefont{{B. Aubert, \emph{et al.}}}(2005)}]{BABAR2005}
\bibinfo{author}{\bibnamefont{{B. Aubert, \emph{et al.}}}}
  (\bibinfo{collaboration}{BABAR Collaboration}), \bibinfo{journal}{Phys. Rev.
  D} \textbf{\bibinfo{volume}{71}}, \bibinfo{pages}{071103}
  (\bibinfo{year}{2005}).

\bibitem[{\citenamefont{{R. Aaij, \emph{et al.}}}(2012)}]{LHCb2012}
\bibinfo{author}{\bibnamefont{{R. Aaij, \emph{et al.}}}}
  (\bibinfo{collaboration}{LHCb Collaboration}), \bibinfo{journal}{Eur. Phys.
  J. C} \textbf{\bibinfo{volume}{72}}, \bibinfo{pages}{1972}
  (\bibinfo{year}{2012}).

\bibitem[{\citenamefont{{Hambrock}}(2013)}]{Hambrock2013}
\bibinfo{author}{\bibfnamefont{C.}~\bibnamefont{{Hambrock}}},
  \bibinfo{journal}{PoS Beauty 044. ArXiv e-prints}  (\bibinfo{year}{2013}),
  \eprint{1306.0695}.

\bibitem[{\citenamefont{{N. A. T\"{o}rnqvist}}(2004)}]{Tornqvist2004}
\bibinfo{author}{\bibnamefont{{N. A. T\"{o}rnqvist}}}, \bibinfo{journal}{Phys.
  Lett B} \textbf{\bibinfo{volume}{590}}, \bibinfo{pages}{209 }
  (\bibinfo{year}{2004}).

\bibitem[{\citenamefont{{ D.~Gamermann, J.~Nieves, E.~Oset and
  E.~R.~Arriola}}(2010)}]{Gamermann:2009uq}
\bibinfo{author}{\bibnamefont{{ D.~Gamermann, J.~Nieves, E.~Oset and
  E.~R.~Arriola}}}, \bibinfo{journal}{Phys. Rev. D}
  \textbf{\bibinfo{volume}{81}}, \bibinfo{pages}{014029}
  (\bibinfo{year}{2010}).

\bibitem[{\citenamefont{{L. Maiani, and F. Piccinini, and A. D. Polosa, and V.
  Riquer}}(2005)}]{Maiani2005}
\bibinfo{author}{\bibnamefont{{L. Maiani, and F. Piccinini, and A. D. Polosa,
  and V. Riquer}}}, \bibinfo{journal}{Phys. Rev. D}
  \textbf{\bibinfo{volume}{71}}, \bibinfo{pages}{014028}
  (\bibinfo{year}{2005}).

\bibitem[{\citenamefont{Barnes and Swanson}(1992)}]{Barnes1992}
\bibinfo{author}{\bibfnamefont{T.}~\bibnamefont{Barnes}} \bibnamefont{and}
  \bibinfo{author}{\bibfnamefont{E.~S.} \bibnamefont{Swanson}},
  \bibinfo{journal}{Phys. Rev. D} \textbf{\bibinfo{volume}{46}},
  \bibinfo{pages}{131} (\bibinfo{year}{1992}).

\bibitem[{\citenamefont{Martins et~al.}(1995)\citenamefont{Martins, Blaschke,
  and Quack}}]{Martins:1994hd}
\bibinfo{author}{\bibfnamefont{K.}~\bibnamefont{Martins}},
  \bibinfo{author}{\bibfnamefont{D.}~\bibnamefont{Blaschke}}, \bibnamefont{and}
  \bibinfo{author}{\bibfnamefont{E.}~\bibnamefont{Quack}},
  \bibinfo{journal}{Phys. Rev. C} \textbf{\bibinfo{volume}{51}},
  \bibinfo{pages}{2723} (\bibinfo{year}{1995}).

\bibitem[{\citenamefont{Pe{\~n}a and Blaschke}(2012)}]{CarlosX}
\bibinfo{author}{\bibfnamefont{C.}~\bibnamefont{Pe{\~n}a}} \bibnamefont{and}
  \bibinfo{author}{\bibfnamefont{D.}~\bibnamefont{Blaschke}},
  \bibinfo{journal}{Acta Phys. Pol. B Proc. Suppl}
  \textbf{\bibinfo{volume}{5}}, \bibinfo{pages}{963} (\bibinfo{year}{2012}).

\bibitem[{\citenamefont{{T.~Barnes, E.~S.~Swanson}}(1992)}]{barnes:1992}
\bibinfo{author}{\bibnamefont{{T.~Barnes, E.~S.~Swanson}}},
  \bibinfo{journal}{Phys. Rev. D} \textbf{\bibinfo{volume}{46}},
  \bibinfo{pages}{131} (\bibinfo{year}{1992}).

\bibitem[{\citenamefont{{T.~Barnes, N.~Black,
  E.~S.~Swanson}}(2001)}]{barnes:2001}
\bibinfo{author}{\bibnamefont{{T.~Barnes, N.~Black, E.~S.~Swanson}}},
  \bibinfo{journal}{Phys. Rev. C} \textbf{\bibinfo{volume}{63}},
  \bibinfo{pages}{025204} (\bibinfo{year}{2001}).

\bibitem[{\citenamefont{Blaschke and R{\"o}pke}(1993)}]{Blaschke:1992qa}
\bibinfo{author}{\bibfnamefont{D.}~\bibnamefont{Blaschke}} \bibnamefont{and}
  \bibinfo{author}{\bibfnamefont{G.}~\bibnamefont{R{\"o}pke}},
  \bibinfo{journal}{Physics Letters B} \textbf{\bibinfo{volume}{299}},
  \bibinfo{pages}{332} (\bibinfo{year}{1993}).

\bibitem[{\citenamefont{{C. Pe{\~n}a}}(2013)}]{CarlosThesis2013}
\bibinfo{author}{\bibnamefont{{C. Pe{\~n}a}}}, \emph{\bibinfo{title}{Quantum
  mechanical model for quarkonium production in heavy ion collisions}}
  (\bibinfo{publisher}{PhD Thesis. University of Wroclaw},
  \bibinfo{address}{Poland}, \bibinfo{year}{2013}).

\bibitem[{\citenamefont{{K. Murasugi}}(1999)}]{MurasugiBraidsBook}
\bibinfo{author}{\bibnamefont{{K. Murasugi}}}, \emph{\bibinfo{title}{Braid
  Groups}} (\bibinfo{publisher}{Kluwer Academic Publishers},
  \bibinfo{year}{1999}).

\bibitem[{\citenamefont{{K. Murasugi}}(1996)}]{Murasugi1996}
\bibinfo{author}{\bibnamefont{{K. Murasugi}}}, \emph{\bibinfo{title}{Knot
  Theory and its Applications}} (\bibinfo{publisher}{{Birkh\"{a}user Boston}},
  \bibinfo{year}{1996}).

\bibitem[{\citenamefont{Artin}(1947)}]{Artin1947}
\bibinfo{author}{\bibfnamefont{E.}~\bibnamefont{Artin}},
  \bibinfo{journal}{Annals of Mathematics, Second Series}
  \textbf{\bibinfo{volume}{48}}, \bibinfo{pages}{101} (\bibinfo{year}{1947}).

\bibitem[{\citenamefont{{C. Kassel, V. Turaev}}(2008)}]{TuraevBook2008}
\bibinfo{author}{\bibnamefont{{C. Kassel, V. Turaev}}},
  \emph{\bibinfo{title}{Braid Groups}} (\bibinfo{publisher}{Springer Science,
  Business Media, LLC}, \bibinfo{year}{2008}).

\bibitem[{\citenamefont{{J. S. Birman}}(1974)}]{Birman1974}
\bibinfo{author}{\bibnamefont{{J. S. Birman}}}, \emph{\bibinfo{title}{Braids,
  Links, and Mapping Class Groups}} (\bibinfo{publisher}{Princeton University
  Press and University of Tokyo Press}, \bibinfo{year}{1974}).

\bibitem[{\citenamefont{{J. Jacak, R. Gonczarek, L. Jacak, I. J\'{o}\'{z}wiak
  }}(2012)}]{Jacak2012}
\bibinfo{author}{\bibnamefont{{J. Jacak, R. Gonczarek, L. Jacak, I.
  J\'{o}\'{z}wiak }}}, \emph{\bibinfo{title}{Application of Braid Groups in 2D
  Hall System Physics Composite Fermion Structure}} (\bibinfo{publisher}{World
  Scientific Publishing Co. Pte. Ltd}, \bibinfo{year}{2012}).

\bibitem[{\citenamefont{{W. Burau}}(1935)}]{Burau1935}
\bibinfo{author}{\bibnamefont{{W. Burau}}}, \bibinfo{journal}{Abhandlungen aus
  dem Mathematischen Seminar der Universit{\"a}t Hamburg}
  \textbf{\bibinfo{volume}{11}}, \bibinfo{pages}{179} (\bibinfo{year}{1935}).

\bibitem[{\citenamefont{{Birman} and {Brendle}}(2004)}]{Birman2004}
\bibinfo{author}{\bibfnamefont{J.~S.} \bibnamefont{{Birman}}} \bibnamefont{and}
  \bibinfo{author}{\bibfnamefont{T.~E.} \bibnamefont{{Brendle}}},
  \bibinfo{journal}{ArXiv e-prints}  (\bibinfo{year}{2004}),
  \eprint{math/0409205}.

\bibitem[{\citenamefont{{N. Jackson}}(2004)}]{Jackson2004}
\bibinfo{author}{\bibnamefont{{N. Jackson}}}, \emph{\bibinfo{title}{Notes on
  braid groups}} (\bibinfo{year}{2004}).

\bibitem[{\citenamefont{{Rolfsen}}(2010)}]{Rolfsen2010}
\bibinfo{author}{\bibfnamefont{D.}~\bibnamefont{{Rolfsen}}},
  \bibinfo{journal}{ArXiv e-prints}  (\bibinfo{year}{2010}),
  \eprint{1010.4051}.

\bibitem[{\citenamefont{{I. N. Bronshtein, K. A. Semendyayev, G. Musiol, H.
  Muehlig}}(2007)}]{Bronshtein2007}
\bibinfo{author}{\bibnamefont{{I. N. Bronshtein, K. A. Semendyayev, G. Musiol,
  H. Muehlig}}}, \emph{\bibinfo{title}{Handbook of Mathematics, 5th Ed}}
  (\bibinfo{publisher}{Springer-Verlag}, \bibinfo{address}{Berlin- Heidelberg},
  \bibinfo{year}{2007}).

\bibitem[{\citenamefont{Hua-Bin and Xiao-Fu}(2014)}]{Zhou2014}
\bibinfo{author}{\bibfnamefont{Z.}~\bibnamefont{Hua-Bin}} \bibnamefont{and}
  \bibinfo{author}{\bibfnamefont{L.}~\bibnamefont{Xiao-Fu}},
  \bibinfo{journal}{Communications in Theoretical Physics}
  \textbf{\bibinfo{volume}{61}}, \bibinfo{pages}{359} (\bibinfo{year}{2014}).

\bibitem[{\citenamefont{{M. Schmidt, G. R\"opke and H.
  Schulz}}(1990)}]{Schmidt199057}
\bibinfo{author}{\bibnamefont{{M. Schmidt, G. R\"opke and H. Schulz}}},
  \bibinfo{journal}{Annals Phys} \textbf{\bibinfo{volume}{202}},
  \bibinfo{pages}{57} (\bibinfo{year}{1990}).

\bibitem[{\citenamefont{Pe{\~n}a and Blaschke}(2014)}]{Carlos2014PRB}
\bibinfo{author}{\bibfnamefont{C.}~\bibnamefont{Pe{\~n}a}} \bibnamefont{and}
  \bibinfo{author}{\bibfnamefont{D.}~\bibnamefont{Blaschke}},
  \bibinfo{journal}{Nuclear Physics A} \textbf{\bibinfo{volume}{927}},
  \bibinfo{pages}{1} (\bibinfo{year}{2014}).

\bibitem[{\citenamefont{Blaschke and Pe{\~n}a}(2011)}]{Carlos2011PRBProc}
\bibinfo{author}{\bibfnamefont{D.}~\bibnamefont{Blaschke}} \bibnamefont{and}
  \bibinfo{author}{\bibfnamefont{C.}~\bibnamefont{Pe{\~n}a}},
  \bibinfo{journal}{Nuclear Physics B - Proceedings Supplements}
  \textbf{\bibinfo{volume}{214}}, \bibinfo{pages}{137} (\bibinfo{year}{2011}).

\end{thebibliography}

\end{document}